\documentclass[sigconf, nonacm]{acmart}

\newcommand\vldbdoi{XX.XX/XXX.XX}

\newcommand\vldbauthors{\authors}
\newcommand\vldbtitle{\shorttitle} 
\newcommand\vldbavailabilityurl{URL_TO_YOUR_ARTIFACTS}
\newcommand\vldbpagestyle{plain} 

\usepackage{tabu}
\usepackage{algorithm}
\usepackage{algpseudocode}
\usepackage{enumitem}
\usepackage{graphicx}
\usepackage{indentfirst}

\begin{document}
\title{Educational Database Prototype: the Simplest of All}

\author{Yi Lyu}
\affiliation{%
  \institution{University of Wisconsin-Madison}
  \city{Madison}
  \state{Wisconsin}
}
\email{ylyu76@wisc.edu}

\author{Yiyin Shen}
\affiliation{%
  \institution{University of Wisconsin-Madison}
  \city{Madison}
  \state{Wisconsin}
}
\email{yshen82@wisc.edu}

\author{Takashi Matsuzawa}
\affiliation{%
  \institution{University of Wisconsin-Madison}
  \city{Madison}
  \state{Wisconsin}
}
\email{takashi@cs.wisc.edu}

\begin{abstract}
\noindent Database Management System (DBMS) is designed to help store and process large collections of data, and is incredibly flexible to perform various kinds of optimizations as long as it achieves serializability with a high-level interface available. The current undergraduate level DBMS course in UW-Madison (i.e., CS564) involves implementing specific modules of DB architecture, including B+ tree, but students may end up spending numerous amounts of effort on corner cases and not gaining a more comprehensive understanding of the internal design. Thus, we present \textit{EduDB}, a simple database prototype for educational purposes that provides students a clean, concise, and comprehensive overview of the database system. We also attempt to develop an integrative series of course projects based on \textit{EduDB}, which offers a platform for students to perform any optimization learned during the semester. 
\end{abstract}

\maketitle

\pagestyle{\vldbpagestyle}
\begingroup\small\noindent\raggedright\textbf{PVLDB Reference Format:}\\
\vldbauthors. \vldbtitle. 
\endgroup

\ifdefempty{\vldbavailabilityurl}{}{
\vspace{.3cm}
\begingroup\small\noindent\raggedright\textbf{PVLDB Artifact Availability:}\\
The source code, data, and/or other artifacts have been made available at \url{https://github.com/yiyins2/Educational-Database-Prototype}. The database is still under development. 
\endgroup
}

\section{Introduction}
\subsection{Motivation}
Database Management System (DBMS) is designed to help store and process large collections of data. Unlike many other Computer Science and Electrical Engineering related fields (e.g., Wireless Networking), DBMS permits researchers and developers to perform a wide range of optimizations while an appropriate high-level interface is presented. While most researches and projects have been working on optimizations for complex commercial databases, we focus on presenting a simple and lightweight database system structure for educational purposes instead. 

\subsubsection{Constructivist-based Learning}
Traditional instructional methods for DBMS courses consist of didactic lectures and a fixed set of assignments. These passive learning methods discourage students from realizing that usually there is no definite implementation of DBMS. Constructivism emphasizes an active learning process in which students construct concepts based on existing foundation to gain new knowledge \cite{connolly2006constructivist}\cite{efflex}\cite{vetrass}\cite{catp}\cite{monom}\cite{zhang2021tapping,zhang2023first,feng2021allign,feng2024f3,feng2025optimus,han2022francis,ctf,safeguard}. Many constructivist-based learning methodologies are proposed, such as problem-based learning and inquiry-based learning. For this project, we want to focus on project-based learning (PBL), in which learners acquire in-depth knowledge through exploring a real-world problem or challenge, ideally out of personal interest. Several studies showed positive feedback of applying PBL in database courses from both the students and the faculties. Connolly's and Begg's study \cite{connolly2006constructivist} showed that students reported that PBL is more motivating and engaging than traditional learning. Faculties felt students acquired more knowledge, especially skills not traditionally covered in database modules. This study also suggests that one integrative project covering several modules would be a powerful instructional approach. Domínguez and Jaime \cite{dominguez2010database} conducted an experiment to test the effectiveness of PBL: students could choose voluntarily to either follow PBL or traditional learning, then the performance (measured by exam dropout rates, exam passing rates, exam marks, and class attendance) of these two student groups were compared. Students who followed PBL obtained better performance, and performance was not mediated by excellent students' preference to follow PBL rather than traditional learning. Naik and Gajjar \cite{naik2021applying} reported that the satisfaction level of both students and faculties is over 90\% for a DBMS class utilizing PBL. Faculties indicated that except for learning the core concepts, students improved their problem-solving skills, confidence, team management, and communication skills. We want to borrow this PBL ideology and implement it in our undergraduate DBMS course at the University of Wisconsin-Madison (CS 564).  

\subsubsection{Limitations of Current Curriculum}
Besides the adoption of PBL, improvements to the current curriculum of DBMS courses are necessary. Saeed \cite{saeed2017role} showed that the current curriculum in most universities fails to provide undergraduate students with the skill to solve poor DB performance in their future careers, and one of the primary reasons is the lack of lower-level practices. To counter this shortcoming, current CS 564 involves implementing specific modules of DB architecture, such as B+ Tree. However, many such assignments turned into busy work for students to cover numerous corner cases. Consequently, students cannot gain a more comprehensive understanding of the DB internal design. Sciore \cite{Boston} proposed that instructors could provide students with the source code of a pedagogically-written DB system and have students modify specific modules. Students can then learn the overall DB structure by studying the source code. In addition, this pedagogically-written DB system should be simple enough to avoid a steep learning curve for students. Its simplicity also permits plenty of room for students to modify the code or add new implementations freely. 

\subsubsection{Goal}
This project implemented an educational DB prototype, named \textit{EduDB}, that consists of an interface that only allows the basic SQL transactions with the minimum amount of coding possible. The objectives of the prototype are the following: 
\begin{enumerate}
    \item It can provide a clean and concise overview of the database system and convey the fact that database system can be extremely simple; 
    \item It provides a platform that students can perform any optimization they have learned throughout the semester (in class or out of class); 
\end{enumerate}
This project also attempts to design an integrative series of experimental projects in which the educational DB prototype serves as a skeleton. Students will be asked to improve the prototype's efficiency by implementing optimizations. They will also participate in a competition, where they are ranked based on the efficiency of the improved prototype against a benchmark. 

\subsection{Related Work}
DBMS-related tools for educational purposes have been introduced early since two decades ago, where Ramakrishnan et al. \cite{2005} proposed a web-based administration tool for projects that may be assigned to students. However, the system itself is mutable; hence, students could only view and modify on a graphic user interface. Guting et al. \cite{20052} proposed a "generic" database system frame for teaching and researching, yet it has similar limitations. 

More recently, many systems have been created for DBMS pedagogical purposes, including: 
\begin{itemize}[leftmargin=2em]
    \item SimpleDB \cite{MIT} is written as course project from MIT during Fall 2010.
    \item SimpleDB \cite{Boston} is developed by Sciore from Boston College. It is written in Java, first proposed in 2007, and latest updated on March 25, 2021. However, this SimpleDB still has over 7,000 lines of codes, which is not simple enough in our view. 
    \item BadgerDB \cite{Badger} is a Database API from UW-Madison. 
\end{itemize}
Apart from educational DB, tinyDB \cite{tiny}, an open source and lightweight database system written in Python, might also be helpful for retrieving a general idea during development processes.

\subsection{Outline}
Section 2 provides a brief architecture overview of \textit{EduDB}, section 3 describes each DB module in details. In section 4, we present the join project we developed and list some other potential projects. Section 5 demonstrates the benchmark and leaderboard we will use to evaluate and rank students' projects. 

\section{Architecture Overview}
\begin{figure}[h]
\includegraphics[width=0.65\linewidth]{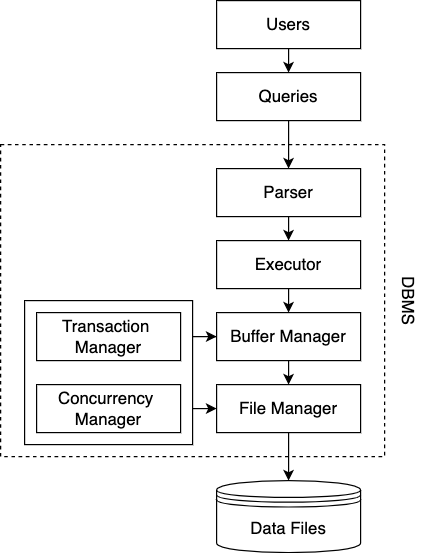}
\caption{Architecture of \textbf{EduDB}}
\label{fig:arch}
\centering
\end{figure}
\textit{EduDB} supports six types of query operations: create table, drop table, insert record, update record, select record, and join tables. The entire architecture to support the above six operations is shown in Figure \ref{fig:arch}. The architecture can be divided into the following modules: 
\begin{itemize}[leftmargin=2em]
    \setlength\itemsep{0.5em}
    \item \textbf{Server-client Mode User Interface.} We use socket to serve as the main communication link between database clients and the database server. The user interface of \textit{EduDB} is terminal-based: users can send requests to and receive query results from the server with terminals.
    \item \textbf{Parser.} The parser contains a set of predefined rules, a sequence of lexers, for each operation. To parse a query, the parser first breaks down the query into lexers and tries to match these incoming lexers with the predefined rules. If a match is found, the query information parsed along the way will be sent to the query executor. Otherwise, the query is invalid. 
    \item \textbf{Query Executor.} The query executor is the central brain of the database. Depends on the parameters generated by the parser, the query executor will trigger corresponding database operation handler and update in-memory metadata. The metadata contains the set of all existing tables' name and the table schema of all existing tables, which speeds up the searching process during query execution.
    \item \textbf{Buffer Manager.} The buffer manager is the "blood" of the database system. It allocates and maintains a buffer pool that contains a fixed number of buffers, which each buffer records the contents of a page linking to a specific block. When there is a pin attempt on a buffer yet the buffer pool is full, the action will wait until there is space or abort after a arbitrary long period of time.
    \item \textbf{File Manager.} The file manager manages all the files and is responsible for creating, reading, and writing the data blocks of a file. In \textit{EduDB}, each table has its own single file. The file manager does all the I/O operations.
    \item \textbf{Concurrency Manager.} The concurrency manager involves a lock table with three basic types of locks: global lock, shared lock and exclusive lock. The global lock is system-size, and is the most straightforward way to achieve serializability. The shared and exclusive lock are table-size. If there is a conflict on a table, the later transaction will wait until the previous transaction unpins the table. If one transaction waits for an arbitrary long time, it will be aborted. No deadlock prevention is implemented. 
    \item \textbf{Transaction Manager.} The transaction manager handles transactions using buffer, file and concurrency manager. When a transaction commits, it flushes and unpins all buffers in the buffer pool, and releases all holding locks. 
    \item \textbf{What we omit in \textit{EduDB}?} For simplicity, we chose not to implement query optimizer, recovery manager and indexing structures. 
\end{itemize}

\begin{table*}[h]
\begin{tabular}{|l|l|l|l|l|l|l|l|l|l|l|l|l|}
\hline
UPDATE & table1 & SET & col1 & = & "str" & WHERE & col2 & = & 2 \\ \hline
keyword & identifier & keyword & identifier & delimiter & constant & keyword & identifier & delimiter & constant \\ \hline
\end{tabular}
\caption{\label{tab:lexer}Lexical breakdown of an update query }
\end{table*}

\section{Detailed Design}
\subsection{General} 
\textit{EduDB} was developed in C++. The parser has about 500 lines of code and the rest has about 1,200 lines of codes. The lines of code are subject to change since the implementation is still under development and systematic testing has not been conducted yet. 
\subsection{Parser} The overall structure of the parser was referred from Sciore's SimpleDB \cite{Boston}. The parser parses valid queries into query information and passes it to the executor, and rejects invalid queries. It has two important helper structures, the Lexer and the Predicate. 
\subsubsection{Lexer} There are four types of lexers: 
\begin{itemize}[leftmargin=2em]
    \item keyword: SQL keywords, such as "select", "from", "where", etc. 
    \item identifier: table name and field name
    \item constant: int or string 
    \item delimiter: asterisk, comma, equal and inequality signs
\end{itemize}

Table \ref{tab:lexer} demonstrates the lexical breakdown of an update command. Each lexer type has a checker function that takes an incoming string and checks whether the string matches the type. For example, the keyword checker checks if the incoming string is in a set of preset keywords, and the int constant checker checks if the string only contains digits. If the match succeeds, the lexer will notify and send the information contained in the string to the parser. Otherwise, the lexer throws an invalid query exception. 

\subsubsection{Predicate}
\begin{figure}[h]
\includegraphics[width=\linewidth]{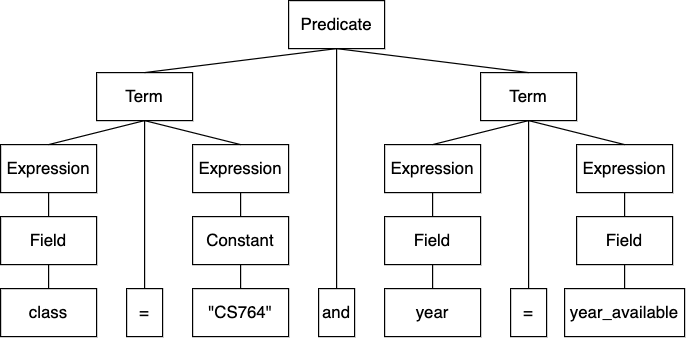}
\caption{Syntax tree of a predicate}
\label{fig:predciate}
\centering
\end{figure}

Each predicate is recursively defined by the following four component: 
\begin{itemize}[leftmargin=2em]
\item predicate: one or more boolean combinations of terms
\item term: a comparison between two expressions
\item expression: constant or field name
\item constant: int or string 
\end{itemize}
Figure \ref{fig:predciate} is a syntax tree of the predicate:
\begin{center}
class = "CS764" and year = year\_available.
\end{center}

\subsubsection{Parser} The parser contains a set of predefined rules for each operation. Each rule is a sequence of lexers. For example, the rule for a create table query is 

\begin{enumerate}
    \item keyword-CREATE 
    \item keyword-TABLE
    \item identifier-table name
    \item delimiter-'('
    \item identifier-field name
    \item keyword-INT or VARCHAR
    \item delimiter-','
    \item repeat 5-7 until no delimiter-',' can be found
    \item delimiter-')' 
\end{enumerate}

The parser breaks down the query into lexers and tries to match these lexers with the predefined rules in the exact order. If a match is not found, then the query is invalid. Otherwise, query information parsed along the way will be passed to the executor. For example, if it is a create table query, the query information will pass to the executor is the table name, field names, and field types.  

\subsection{Query Executor}
\begin{figure}[h]
\includegraphics[width=0.8\linewidth]{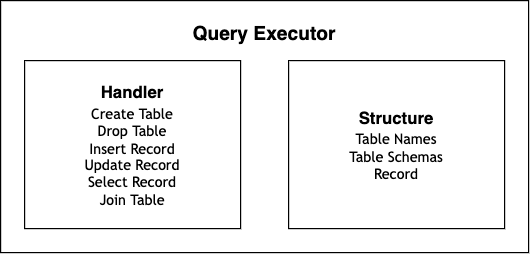}
\caption{\textit{EduDB}'s query executor}
\label{fig:queryExecutor}
\centering
\end{figure}

The query executor is mainly responsible for executing the database operations and maintaining database metadata. The structure of query executor is shown in Figure \ref{fig:queryExecutor}. Given the parser's result, the query executor triggers the corresponding handler.

Since our goal is to implement a simple database protocol for the student to improve on, we do not keep in-memory data structure to help identify the location of the records. If we want to find out whether some records exist in some table, we simply read the whole file of the table. How the query executor support the six operations is listed below: 
\begin{itemize}[leftmargin=2em]
    \setlength\itemsep{0.5em}
    \item \textbf{Create Table.} When the client wants to create a table, the query executor will get the table name with specified field names generated by parser. Since each data record shares the same table schema, the query executor will store the mapping of table name and column fields in memory. The content in table's data file are all binary bytes. We can only recover all data if we know the table schema (the size of data record and each column's meaning).
    \item \textbf{Drop Table.} Dropping (Deleting) a table is easy. The query executor will remove the table's binary file and corresponding information from the metadata.
    \item \textbf{Insert Record.} In \textit{EduDB}, we do not have default value for empty fields. So new records must contain values for all predefined fields. Then the file manager will append the new records to the end of that table's data file. 
    \item \textbf{Update Record.} The file manager will iterate all records of the given table and check whether they satisfy given predicates. If they do, the file manager will update its corresponding fields.
    \item \textbf{Select Record.} The file manager will iterate all records of the given table and check whether they satisfy given predicates. If they do, the file manager will add these records into their result set.
    \item \textbf{Join Tables.} The detailed join algorithm is shown in Algorithm \ref{alg:cap}. The file manager will iterate the cross product of all records from the specified tables. If they satisfy the result, add corresponding fields into result set.
    
    \vspace{-0.5\baselineskip}
    \begin{minipage}{\linewidth}
    \begin{algorithm}[H]
    \caption{Join Algorithm of \textit{EduDB}}\label{alg:cap}
    \begin{algorithmic}
    \State \textbf{Input:} $T_{A}, T_{B}, Pred, F()$
    \State \textbf{Output: S}
    \For{$R_{A}$ in $T_A$}
    \For{$R_{B}$ in ${T_B}$}
        \If{$F(R_{A}, R_{B}, Pred)$}
            \State \textbf{S} $\leftarrow S \lor p_{A}p_{B}$
    \EndIf
    \EndFor
    \EndFor
    \end{algorithmic}
    \end{algorithm}
    \end{minipage}
    \vspace{0.5\baselineskip}
    
    Here $T_A$ and $T_B$ are the two tables that are to be joined. $Pred$ is the join condition specified from the user's request. $F()$ is the function to tell whether the records satisfy the predicate. $R_A$ and $R_B$ are the records from Table A and Table B. $p_A$ and $p_B$ are the projections of the specified table columns.
\end{itemize}

\subsection{Buffer Manager}
When a buffer manager is created, it allocates a buffer pool with a fixed size of buffers. A buffer tracks the contents of a page, its corresponding data block id, and number of pins on it. Whenever a buffer is assigned to a new page, it flushes the dirty data and reads the page using file manager, and then resets the number of pins to one.

\begin{figure}[th]
\includegraphics[width=0.45\linewidth]{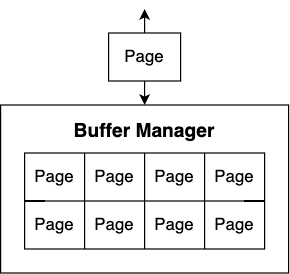}
\caption{\textit{EduDB}'s buffer manger}
\label{fig:buffermanager}
\centering
\end{figure}

When a transaction attempts to pin a new block yet the buffer pool is already full, a conditional variable is called to wait and check periodically (one second by default). If there is still no buffer available after arbitrary long period of time (ten seconds by default), the transaction is aborted with message "Buffer Manager Abort: not enough space". As soon as a buffer is unpinned and becomes available, the conditional variable notifies the waiting pin attempt if any. Pseudo-code is shown in Algorithm \ref{alg:pin}. The pin and unpin actions are protected by mutex to prevent inconsistent results.

\begin{algorithm}[H]
\caption{Algorithm of pin action in buffer manager}\label{alg:pin}
\begin{algorithmic}
\State \textbf{Input:} $block$  
\If{$buf$ linked with $block$ in $pool$}
    \State $pin$ ($buf$)
\ElsIf{$free\_buf$ in $pool$}
    \State $free\_buf \rightarrow link\_block$($block$)
    \State $pin$ ($free\_buf$)
\Else
    \While{not wait too long and $free\_buf$ not in $pool$}
        \State $cv \rightarrow wait$(short time)
    \EndWhile
    \If{$free\_buf$ in $pool$}
        \State $free\_buf \rightarrow link\_block$($block$)
        \State $pin$ ($free\_buf$)
    \Else
        \State $Abort$("Buffer Manager Abort: not enough space")
    \EndIf
\EndIf
\end{algorithmic}
\end{algorithm}

\subsection{File Manager}
File manager is used for reading and writing the specific data block of certain data file (table). File manager will only be called when the page is missed in the buffer manager. Here all data fields are stored in binary format to make the data size consistent.

\begin{figure}[th]
\includegraphics[width=0.45\linewidth]{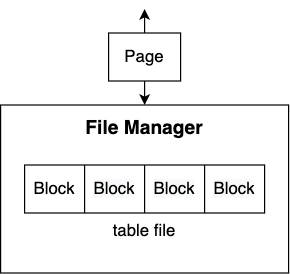}
\caption{\textit{EduDB}'s file manger}
\label{fig:filemanager}
\centering
\end{figure}

\subsection{Concurrency Manager}
Concurrency manager is used for concurrency controls among multiple transactions. \textit{EduDB} involves three basic types of lock modes:

\begin{itemize}[leftmargin=2em]
\item \textbf{Global lock:} implemented for simplicity purpose of the project. It holds a system-size lock and thus during meantime no other action is allowed to perform.
\item \textbf{Shared lock:} the read lock used for reading actions only. It passes a table as parameter and holds the lock upon. Shared lock is compatible with other shared locks, but not global lock or exclusive locks upon the same block.
\item \textbf{Exclusive lock:} the write lock used for writing actions only. Like shared lock, it also passes a table as parameter and holds the lock upon. However, exclusive lock is not compatible with global lock or other shared or exclusive locks.
\end{itemize}
\begin{table}[th]
    \begin{tabular}{|c|c|c|c|}
    \hline
     & Global & Shared & Exclusive  \\ \hline
    Global & False & False & False \\ \hline
    Shared & False & True & False  \\ \hline
    Exclusive & False & False & False \\ \hline
    \end{tabular}
    \caption{\label{tab:lock table}Lock compatibility matrix }
\end{table}

Every successful lock action calls a conditional variable. When a transaction attempts to grant a lock but encounter a conflict with another currently holding lock, it is assigned to a wait list. As the previous lock get released, the conditional variable notifies the wait list, and the transaction can now hold the first lock with the corresponding table. Due to simplicity purpose, no deadlock detection nor prevention is implemented.

\subsection{Transaction Manager}
The idea of transaction manager is simple. Each transaction has its own unique integer as identifier. The number starts with zero and increment by one for every new transaction. The transaction passes a file manager and buffer manager as parameters, and initialize a concurrency manager during construction. When the transaction commits, it unpins all buffers in the buffer pool and release all currently holding locks (global locks by default for simplicity).

\section{Course Projects}
As mentioned in our motivation, we would like to design an integrative series of projects to help students understand the DB internals and encourage them to practice implementing their own optimizations. We already implemented the join project and are considering adding more projects for future work. 

\subsection{The Join Project}
Current implementation of joining tables is simple nested join, see section 3.3. Students can take the liberty of implementing any hash join algorithm.

We propose a plausible grading scheme for the join project: TAs provide auto-tests that execute queries involving join. Students who have finished their codes can benchmark their versions on an automated server and compare the results (i.e. elapsed time) with the nested join example. More details of the benchmark and the leaderboard can be found in Section 5.

Feedback can be presented in a predetermined format. For example, "In the test X, the submission is X percents faster/slower than nested loop join." After students finalizing their submissions, TAs only need to manually review the ones that fail to exceed certain threshold, thus alleviate their grading workloads. 

\subsection{Other Potential Projects}
Aside from the join project, the following projects are plausible to develop for future work: 
\begin{itemize}[leftmargin=2em]
\item \textbf{Optimized Concurrency Manager.} Since buffer pool and shared \& exclusive locks might not be essential for a database system to function (naive buffers and global lock might be sufficient), these two features can be used as a potential course project. Students will be asked to implement the buffer pool, locks on a finer granularity (e.g., block) and some deadlock prevention mechanism (e.g., wait-die). 
\item \textbf{Indexing.} Students can implement different indexing structure, such as B-Trees, hash tables, and bloom filters; for range, point, and existence queries. 
\end{itemize}

\section{Benchmark and Leaderboard}
Since students' working machines have different hardware environments, it is more reliable to benchmark their code on one designated server machine. However, the testing result can also be influenced by the server's workload at that time. Generally speaking, the same code will get a better result if the total workload of the server is low. Also, if all the students are running the codes simultaneously, it may overwhelm our server. Thus it should be the server who autonomously runs the benchmark for the students.

\begin{figure}[th]
\includegraphics[width=\linewidth]{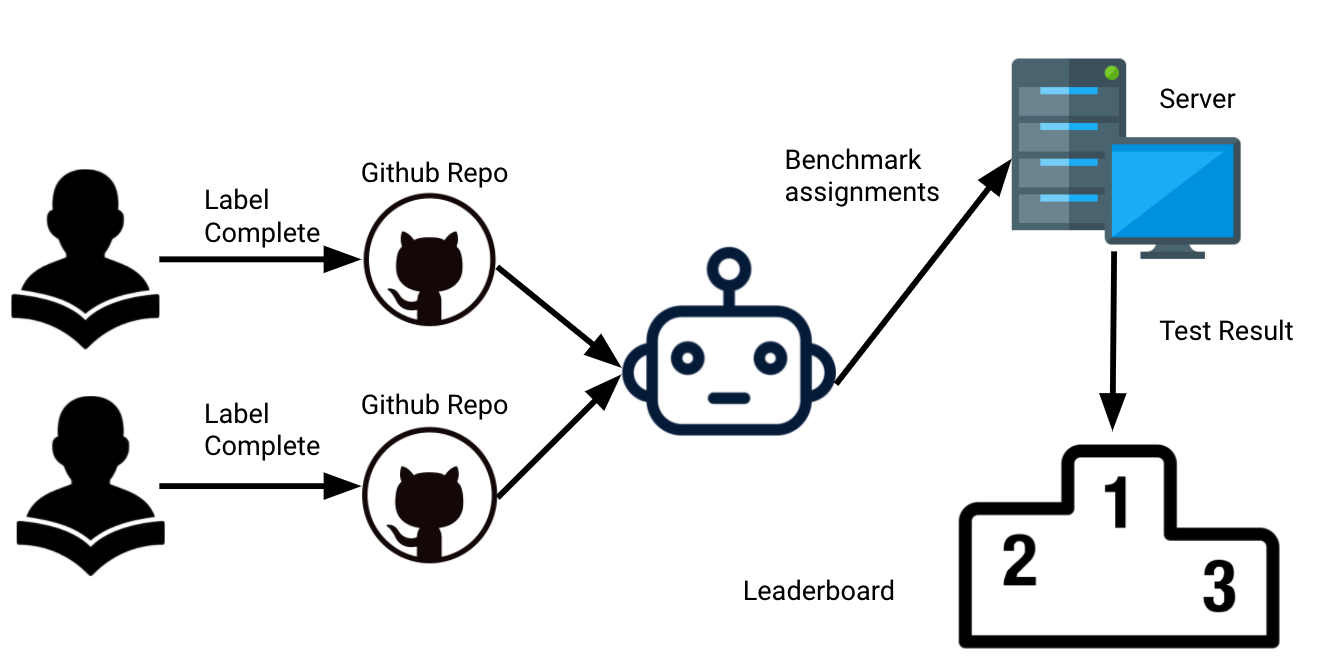}
\caption{\textit{EduDB}'s benchmark}
\label{fig:benchmark}
\centering
\end{figure}

Our current strategy is shown in Graph \ref{fig:benchmark}. First, each student will have a GitHub repository submitting their codes. A bot code will then automatically detect whose repositories are labeled as complete. Periodically the bot will run the students' codes on the test server. We can modify the parameters to control the testing scale and frequency to provide a comparably reliable and stable testing environment.

\begin{figure}[h]
\includegraphics[width=\linewidth]{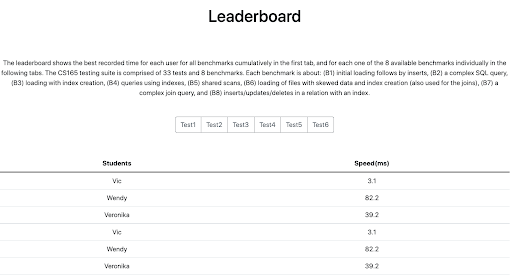}
\caption{\textit{EduDB}'s Leader board}
\label{fig:leaderboard}
\centering
\end{figure}

Finally, we created a leader board website written in Django and React, which will show the result and ranking of the students on the website. The demo is shown in Figure \ref{fig:leaderboard}.

\section{Future Work}
\textit{EduDB} are currently lacking:
\begin{itemize}[leftmargin=2em]
    \item \textbf{Systematic large-scale testing.} Currently we have several small-scale and simple test cases for the server, the parser, and the concurrency manager, yet systematic and large-scale tests are also required to test the functionality and serializability of the system.
    \item \textbf{Commenting and documentation.} Although the system is currently executable, additional comments, instructions and documentations are needed to be usable by the students in the future. 
    \item \textbf{Course projects.} More details of the projects needs to be implemented, including the instructions, test cases, and possible hints.
    \item \textbf{Join query.} Other query operations are mostly finished and executable, but the join operation is still having some issues and need to be fixed. 
\end{itemize}

\section{Conclusion}
In this project, we presented \textit{EduDB}, a simple database prototype for educational purposes. We explained the details of the parser, query executor, buffer manager, file manager, concurrency manager, and transaction manager. Query optimize, recovery manager and indexing structures were omitted for simplicity. We also proposed the Join project and listed other potential course projects. Lastly, we introduced a generic workflow of \textit{EduDB}'s benchmark to evaluate and rank students' optimizations. 

Our main contribution was attempting to design and construct the simplest database system that is executable and serializable and provide a platform for students to implement any optimization learned during or outside the course. We hope the prototype can be used as a foundation of the future course materials in CS564.

\bibliographystyle{ACM-Reference-Format}
\bibliography{sample}

@String{Computing = "Computing" }

@String{Computer = "{IEEE} Computer" }

@Article{saeed2017role,
  title={Role of database management systems (dbms) in supporting information technology in sector of education},
  author={Saeed, Adham Mohsin},
  journal={Int. J. Sci. Res},
  volume={6},
  number={5},
  pages={1462--1466},
  year={2017}
}

@article{2005,
author = {Ramakrishnan, Sub and Nwosu, Emeka},
title = {DBMS Course: Web Based Database Administration Tool and Class Projects},
year = {2003},
issue_date = {January 2003},
publisher = {Association for Computing Machinery},
address = {New York, NY, USA},
volume = {35},
number = {1},
issn = {0097-8418},
url = {https://doi.org/10.1145/792548.611922},
doi = {10.1145/792548.611922},
journal = {SIGCSE Bull.},
month = jan,
pages = {16–20},
numpages = {5}
}

@INPROCEEDINGS{20052,
  author={Guting, R.H. and Almeida, V. and Ansorge, D. and Behr, T. and Ding, Z. and Hose, T. and Hoffmann, F. and Spiekermann, M. and Telle, U.},
  booktitle={21st International Conference on Data Engineering (ICDE'05)}, 
  title={SECONDO: an extensible DBMS platform for research prototyping and teaching}, 
  year={2005},
  volume={},
  number={},
  pages={1115-1116},
  doi={10.1109/ICDE.2005.129}}

@article{MIT,
  author={Madden, Samuel and Morris, Robert and Stonebraker, Michael and Curino, Carlo},
  title={6.830 Database Systems. Fall 2010. Massachusetts Institute of Technology: MIT OpenCourseWare https://ocw.mit.edu. License: Creative Commons BY-NC-SA.}, 
  year={2020}}

@inproceedings{Boston,
author = {Sciore, Edward},
title = {SimpleDB: A Simple Java-Based Multiuser Syst for Teaching Database Internals},
year = {2007},
isbn = {1595933611},
publisher = {Association for Computing Machinery},
address = {New York, NY, USA},
url = {https://doi.org/10.1145/1227310.1227498},
doi = {10.1145/1227310.1227498},
booktitle = {Proceedings of the 38th SIGCSE Technical Symposium on Computer Science Education},
pages = {561–565},
numpages = {5},
location = {Covington, Kentucky, USA},
series = {SIGCSE '07}
}

@article{naik2021applying,
  title={Applying and Evaluating Engagement and Application-Based Learning and Education (ENABLE): A Student-Centered Learning Pedagogy for the Course Database Management System},
  author={Naik, Shefali and Gajjar, Kunjal},
  journal={Journal of Education},
  pages={00220574211032319},
  year={2021},
  publisher={SAGE Publications Sage CA: Los Angeles, CA}
}

@article{dominguez2010database,
  title={Database design learning: A project-based approach organized through a course management system},
  author={Dom{\'\i}nguez, C{\'e}sar and Jaime, Arturo},
  journal={Computers \& Education},
  volume={55},
  number={3},
  pages={1312--1320},
  year={2010},
  publisher={Elsevier}
}

@website{Badger,
  author={Haiyun Jin},
  title={BadgerDB},
  url={http://pages.cs.wisc.edu/~jignesh/cs564/projects/BadgerDB/BufMgr/docs/}}

@website{tiny,
  title={tinyDB},
  year={2021},
  url={https://tinydb.readthedocs.io/en/latest/}}

@article{connolly2006constructivist,
  title={A Constructivist-Based Approach to Teaching Database Analysis and Design.},
  author={Connolly, Thomas M and Begg, Carolyn E},
  journal={Journal of Information Systems Education},
  volume={17},
  number={1},
  year={2006},
  publisher={Citeseer}
}

@InProceedings{efflex,
    author    = {Cheng, Ming and Zhou, Ziyi and Zhang, Bowen and Wang, Ziyu and Gan, Jiaqi and Ren, Ziang and Feng, Weiqi and Lyu, Yi and Zhang, Hefan and Diao, Xingjian},
    title     = {Efflex: Efficient and Flexible Pipeline for Spatio-Temporal Trajectory Graph Modeling and Representation Learning},
    booktitle = {Proceedings of the IEEE/CVF Conference on Computer Vision and Pattern Recognition (CVPR) Workshops},
    month     = {June},
    year      = {2024},
    pages     = {2546-2555}
}

@misc{vetrass,
      title={VeTraSS: Vehicle Trajectory Similarity Search Through Graph Modeling and Representation Learning}, 
      author={Ming Cheng and Bowen Zhang and Ziyu Wang and Ziyi Zhou and Weiqi Feng and Yi Lyu and Xingjian Diao},
      year={2024},
      eprint={2404.08021},
      archivePrefix={arXiv},
      primaryClass={cs.LG},
      url={https://arxiv.org/abs/2404.08021}, 
}

@INPROCEEDINGS{catp,
  author={Liao, Ruqi and Zhao, Chuqing and Li, Jin and Feng, Weiqi and Lyu, Yi and Chen, Bingxian and Yang, Haochen},
  booktitle={2025 IEEE Conference on Artificial Intelligence (CAI)}, 
  title={CATP: Cross-Attention Token Pruning for Accuracy Preserved Multimodal Model Inference}, 
  year={2025},
  volume={},
  number={},
  pages={1100-1104},
  doi={10.1109/CAI64502.2025.00191}}

@misc{monom,
      title={MonoM: Enhancing Monotonicity in Learned Cardinality Estimators}, 
      author={Lyu Yi and Weiqi Feng and Yuanbiao Wang and Yuhong Kan},
      year={2025},
      eprint={2512.22122},
      archivePrefix={arXiv},
      primaryClass={cs.DB},
      url={https://arxiv.org/abs/2512.22122}, 
}

@article{zhang2021tapping,
  title={Tapping into nfv environment for opportunistic serverless edge function deployment},
  author={Zhang, Lu and Feng, Weiqi and Li, Chao and Hou, Xiaofeng and Wang, Pengyu and Wang, Jing and Guo, Minyi},
  journal={IEEE Transactions on Computers},
  volume={71},
  number={10},
  pages={2698--2704},
  year={2021},
  publisher={IEEE}
}

@inproceedings{zhang2023first,
  title={First: Exploiting the multi-dimensional attributes of functions for power-aware serverless computing},
  author={Zhang, Lu and Li, Chao and Wang, Xinkai and Feng, Weiqi and Yu, Zheng and Chen, Quan and Leng, Jingwen and Guo, Minyi and Yang, Pu and Yue, Shang},
  booktitle={2023 IEEE International Parallel and Distributed Processing Symposium (IPDPS)},
  pages={864--874},
  year={2023},
  organization={IEEE}
}

@inproceedings{feng2021allign,
  title={Allign: Aligning all-pair near-duplicate passages in long texts},
  author={Feng, Weiqi and Deng, Dong},
  booktitle={Proceedings of the 2021 International Conference on Management of Data},
  pages={541--553},
  year={2021}
}

@inproceedings{feng2025optimus,
  title={Optimus: Accelerating $\{$Large-Scale$\}$$\{$Multi-Modal$\}$$\{$LLM$\}$ Training by Bubble Exploitation},
  author={Feng, Weiqi and Chen, Yangrui and Wang, Shaoyu and Peng, Yanghua and Lin, Haibin and Yu, Minlan},
  booktitle={2025 USENIX Annual Technical Conference (USENIX ATC 25)},
  pages={161--177},
  year={2025}
}

@article{feng2024f3,
  title={F3: Fast and Flexible Network Telemetry with an FPGA coprocessor},
  author={Feng, Weiqi and Gao, Jiaqi and Chen, Xiaoqi and Antichi, Gianni and Basat, Ran Ben and Shao, Michael Mingchao and Zhang, Ying and Yu, Minlan},
  journal={Proceedings of the ACM on Networking},
  volume={2},
  number={CoNEXT4},
  pages={1--22},
  year={2024},
  publisher={ACM New York, NY, USA}
}

@article{han2022francis,
  title={Francis: Fast reaction algorithms for network coordination in switches},
  author={Han, Wenchen and Feng, Vic and Schwartzman, Gregory and Li, Yuliang and Mitzenmacher, Michael and Yu, Minlan and Ben-Basat, Ran},
  journal={arXiv preprint arXiv:2204.14138},
  year={2022}
}

@misc{ctf,
      title={CTF for education}, 
      author={Yi Lyu and Luke Dotson and Nic Draves and Andy Zhang},
      year={2026},
      eprint={2601.17543},
      archivePrefix={arXiv},
      primaryClass={cs.CR},
      url={https://arxiv.org/abs/2601.17543}, 
}

@misc{safeguard,
      title={Safeguard: Security Controls at the Software Defined Network Layer}, 
      author={Yi Lyu and Shichun Yu and Joe Catudal},
      year={2026},
      eprint={2601.17355},
      archivePrefix={arXiv},
      primaryClass={cs.CR},
      url={https://arxiv.org/abs/2601.17355}, 
}

\end{document}